\documentstyle[11pt,cs10]{article}

\makeatletter
\def\astropbibitem{\@lbibitem}
\def\@lbibitem#1#2#3{\item[\hfill]\if@filesw 
      { \def\protect##1{\string ##1\space}\immediate
        \write\@auxout{\string\astropbibcite{#3}{#1}{#2}}\fi\ignorespaces}}
\def\astropbibcite#1#2#3{\global\@namedef{b@#1}{#2\ #3} %without comma 
			\global\@namedef{newb@#1}{#2\ (#3}
			\global\@namedef{nameb@#1}{#2}
			\global\@namedef{yearb@#1}{#3}}
\let\citation\@gobble

\def\cite{\@ifnextchar [{\@tempswatrue\@citex}{\@tempswafalse\@citex[]}}
\def\citeone{\@ifnextchar [{\@tempswatrue\@newcitex}
			   {\@tempswafalse\@newcitex[]}}
\def\citename{\@ifnextchar [{\@tempswatrue\@namecitex}
			   {\@tempswafalse\@namecitex[]}}
\def\citeyear{\@ifnextchar [{\@tempswatrue\@yearcitex}
			   {\@tempswafalse\@yearcitex[]}}
\def\@citex[#1]#2{\if@filesw\immediate\write\@auxout{\string\citation{#2}}\fi
  \def\@citea{}\@cite{\@for\@citeb:=#2\do
     {\@citea\def\@citea{; }\@ifundefined
       {b@\@citeb}{{\bf ?}\@warning
       {Citation `\@citeb' on page \thepage \space undefined}}%
{\csname b@\@citeb\endcsname}}}{#1}}
\def\@newcitex[#1]#2{\if@filesw\immediate\write\@auxout{\string\citation{#2}}\fi
  \def\@newcitea{}\@newcite{\@for\@newciteb:=#2\do
     {\@newcitea\def\@newcitea{; }\@ifundefined
       {newb@\@newciteb}{{\bf ? (?}\@warning
       {Citation `\@newciteb' on page \thepage \space undefined}}%
{\csname newb@\@newciteb\endcsname}}}{#1}}

\def\@namecitex[#1]#2{\if@filesw\immediate\write\@auxout{\string\citation{#2}}\fi
  \def\@namecitea{}\@namecite{\@for\@nameciteb:=#2\do
     {\@namecitea\def\@namecitea{; }\@ifundefined
       {nameb@\@nameciteb}{{\bf ? (?}\@warning
       {Citation `\@nameciteb' on page \thepage \space undefined}}%
{\csname nameb@\@nameciteb\endcsname}}}{#1}}

\def\@yearcitex[#1]#2{\if@filesw\immediate\write\@auxout{\string\citation{#2}}\fi
  \def\@yearcitea{}\@yearcite{\@for\@yearciteb:=#2\do
     {\@yearcitea\def\@yearcitea{; }\@ifundefined
       {yearb@\@yearciteb}{{\bf ? (?}\@warning
       {Citation `\@yearciteb' on page \thepage \space undefined}}%
{\csname yearb@\@yearciteb\endcsname}}}{#1}}

\let\bibdata=\@gobble
\let\bibstyle=\@gobble

\def\bibliography#1{\if@filesw\immediate\write\@auxout{\string\bibdata{#1}}\fi
  \@input{\jobname.bbl}}

\def\bibliographystyle#1{\if@filesw\immediate\write\@auxout
    {\string\bibstyle{#1}}\fi}
\def\nocite#1{\@bsphack
  \if@filesw\immediate\write\@auxout{\string\citation{#1}}\fi
  \@esphack}
\def\@cite#1#2{({#1\if@tempswa ; #2\fi})}
\def\@newcite#1#2{{#1\if@tempswa ; #2\fi})}
\def\@namecite#1#2{#1}
\def\@yearcite#1#2{#1}
\newcommand{\citebare}[1]{{\citename{#1}\ \citeyear{#1}}} %without comma 
\newcommand{\citeeg}[1]{{(e.g., {}\citebare{#1})}}

\def\thebibliography#1{\section*{References}\list
 {}{\setlength\labelwidth{1.4em}\leftmargin\labelwidth
 \setlength\parsep{0pt}\setlength\itemsep{0pt}
 \setlength{\itemindent}{-\leftmargin}
 \usecounter{enumi}}
 \def\newblock{\hskip .11em plus .33em minus -.07em}
 \sloppy
 \sfcode`\.=1000\relax}

\makeatother

\newcommand{\eBoo}{\mbox{$\eta$~Boo}}
\makeatletter 
 \def\sub#1{\relax\ifmmode _{\fam\z@ #1}\else
         $_{\fam\z@ #1}$\fi}
 \def\super#1{\relax\ifmmode ^{\fam\z@ #1}\else
         $^{\fam\z@ #1}$\fi}
\makeatother 
\newcommand{\down}[2]{#1\sub{#2}}
\newcommand{\Teff}{\down{T}{eff}}
\newcommand{\aCenA}{\mbox{$\alpha$~Cen~A}}

\begin{document}

\title{Solar-like oscillations: the search goes on}

\author{Timothy R. Bedding}
\affil{School of Physics, University of Sydney 2006, Australia}

\author{Hans Kjeldsen} 

\affil{Teoretisk Astrofysik Center, Danmarks Grundforskningsfond, Aarhus
University, DK-8000 Aarhus~C, Denmark}

\begin{abstract}

Despite numerous attempts and many hours of telescope time, there has so
far been no confirmed detection of solar-like oscillations in any star
except the Sun.  We review recent efforts, with particular emphasis on the
technique of monitoring equivalent widths of Balmer lines and the steps in
data reduction.

\end{abstract}


\section{Introduction}

Measuring stellar oscillations is a beautiful physics experiment.  A star
is a gaseous sphere and will oscillate in one or more modes when suitably
excited.  The best targets are stars which oscillate in several modes
simultaneously.  Each mode has a slightly different frequency, reflecting
spatial variations of the sound speed within the star, which in turn
depends on density, temperature, gas motion and other properties of the
stellar interior.  The oscillation amplitudes are determined by the
excitation and damping processes, which may involve opacity variations,
turbulence from convection and magnetic fields.  Studying the frequencies
and amplitudes of oscillations in different types of stars promises to lead
to significant advances in our understanding of stellar structure and
evolution (for recent reviews see \citebare{B+G94}; \citebare{G+S96}).

The best-studied example of an oscillating star is the Sun.  Observations
of the 5-minute solar oscillations have led to enormous progress in our
understanding of solar and stellar theory \cite{G+T91} and it is widely
expected that measuring oscillation frequencies in other Sun-like stars
will produce similar advances.  Oscillations in the Sun are excited by
convective turbulence near the surface, so all stars with an outer
convective zone should undergo similar oscillations.  This makes it
possible, at least in principle, to perform seismic studies on {\em all\/}
stars with spectral type later than about~F5.  For our purposes, we define
solar-like oscillations to be those which are excited stochastically by
convection.

%Accurate oscillation data will produce a confrontation with the theory of
%stellar structure and allow us to assess the importance of phenomena that
%are not included in traditional treatments of stellar evolution.

An advantage of studying solar-like oscillations is that the modes are easy
to identify.  There is little point in knowing the frequency of an
oscillation mode unless you also know in which part of the star that mode
is trapped.  An oscillation mode is characterized by three integers:
$n$~(the radial order), $\ell$~(the angular degree) and $m$~(the azimuthal
order)\footnote{In a star with no rotation or magnetic field, frequencies
do not depend on~$m$.}.  These specify the shape of the eigenfunction,
which in turn determines the sensitivity of the oscillation frequency to
the internal structure of the star.  In the Sun, as opposed to classical
variables (e.g., $\delta$~Scuti stars, rapidly oscillating Ap stars and
$\beta$~Cephei stars), all modes in a broad frequency range are excited.
Furthermore, these modes approximately satisfy an asymptotic relation, with
modes of fixed $\ell$ and differing $n$ having regularly spaced frequencies
separated by the so-called large separation,~$\Delta\nu$.  The resulting
comb-like structure allows modes to be identified directly from the
oscillation spectrum.

Measuring $\Delta\nu$ provides an estimate of the stellar density.
Moreover, the small differences between observed frequencies and those
predicted by the asymptotic relation give crucial information about the
sound speed deep inside the star.  For example, in the Sun we find that
modes with $\ell=2$ are displaced by a few per cent of $\Delta\nu$ from
modes with $\ell=0$.  This displacement contains information on the
internal properties of the Sun, such as helium content.

\section{Detection methods}

The disadvantage of studying solar-like oscillations is their tiny
amplitudes.  Three methods have been tried:

\paragraph{Velocity} 

In the Sun, the strongest modes have velocity amplitudes of about 25\,cm/s,
which corresponds to a wavelength variation ($\delta\lambda/\lambda$) of
less than $10^{-9}$, or 4.2\,$\mu$\AA{} at 5000\,\AA\@.  Detecting such
miniscule Doppler shifts in other stars is extremely difficult.
Spectrographs cannot be made with absolute stabilities of $10^{-9}$, so one
must simultaneously monitor the wavelength of a stable reference (e.g., a
Na or K resonance cell, an $I_{2}$ absorption cell or telluric absorption
features).  The noise levels at present are down to about 0.5\,m/s, which
is a factor of two higher than the solar signal.

\paragraph{Intensity} 

The solar oscillations have been observed as variations in total intensity,
with amplitudes of about 4~ppm (parts per million).  Open clusters are a
natural target for differential CCD photometry and the lowest noise level
so far achieved is 5--7 ppm, from observations by \citeone{GBK93} of twelve
stars in M\,67 using six telescopes (2.5\,m to 5\,m) during one week.  This
is an interesting noise level, less than a factor of two away from solar
photometric amplitude.

Ground-based photometric observations are severely hampered by atmospheric
scintillation.  Several space missions have been proposed, but only one has
so far been launched: the EVRIS experiment, on board the Russian Mars96
probe, which ended in the Pacific Ocean.

\paragraph{Temperature} 

Since the change in radius during solar oscillations is insignificant, the
intensity fluctuations observed in the Sun must result from local
temperature changes in the atmosphere of about 6\,mK ($\delta\Teff/\Teff
\approx 10^{-6}$).  It has been suggested that these temperature changes
can be measured by their effect on spectral absorption lines
\cite{KBV95,BKR96}.  For example, the Balmer lines in the Sun should show
variations in equivalent width of about 6\,ppm.  As discussed below, the
equivalent-width method has so far attained noise levels in other stars of
2--3 times the solar peak amplitude (and even less for \aCenA).

\section{Recent results}

There have been many unsuccessful attempts over the past decade to measure
oscillations in other solar-like stars.  This continuing commitment
reflects both the extreme difficulty of the observations and the tremendous
importance that is attached to a successful result \citeeg{B+G94}.  Indeed,
it is fair to say that theorists have been waiting eagerly -- and with some
frustration -- for the first oscillation data to appear.  Attempts to
detect solar-like oscillations have been reviewed by \citeone{B+G94} and
\citeone{K+B95}, and here we only discuss more recent results.  Most
efforts have concentrated on subgiants, since these are expected to have
higher oscillations amplitudes than the Sun.

\paragraph{\eBoo} This star is the brightest G-type subgiant.  We observed
\eBoo{} over six nights with the 2.5-m Nordic Optical Telescope
\cite{KBV95,B+K95}.  Using the equivalent-width method, we claimed a
detection of solar-like oscillations with amplitudes at the expected level
and frequencies that were subsequently shown to be consistent with models
(Christensen-Dalsgaard, Bedding \& Kjeldsen \citeyear{ChDBK95},
\citebare{Gu+D96}).  Since then, the improved luminosity estimate from
Hipparcos measurements has given even better agreement \cite{BKChD98}.

However, a search for velocity oscillations in \eBoo{} by \citeone{BKK97}
has failed to detect a signal, setting limits at a level below the value
expected on the basis of the \citename{KBV95} result.  Brown et al.\
(private communication) have a more recent and larger set of observations
which they are currently processing.

\paragraph{The Sun}

Some support for the equivalent-width method was given by \citeone{KHB97},
who detected the 5-minute oscillations in the Sun from spatially resolved
measurements of H$\beta$ equivalent widths.

\paragraph{\aCenA} This is the brightest G-type main-sequence star.  We
obtained H$\alpha$ spectra over six nights in April 1995 using the 3.9-m
AAT (UCLES) and the 3.6-m ESO (CASPEC).  The observations were done in
collaboration with S. R. Frandsen and T. H. Dall (Aarhus Univ.).  Data
reduction using the equivalent-width method was hampered by a variability
of the continuum, which seems to be due to some kind of colour term in
scintillation at a level of about $10^{-4}$ per minute (well below the
normal photometric scintillation).  Oscillations were not detected, with an
upper limit only slightly higher than the expected signal (Kjeldsen et al.,
in preparation).

\paragraph{Procyon} 
This star is the brightest F-type subgiant in the sky.  Recent results from
Doppler-shift measurements are: (i)~\citeone{BCC95}, using a narrow-band
filter, have retracted an earlier possible detection; and
(ii)~\citeone{BKN96}, using an echelle spectrograph, have not detected a
signal.  More recently, Brown et al.\ (poster paper at this conference)
have obtained new measurements which appear to confirm the excess of power
found previously \cite{BGN91}.

We observed Procyon for several hours per night during the April 1995 run
mentioned above.  Preliminary analysis of Balmer line equivalent widths
appeared to show excess power at the expected amplitude and frequency, but
we no longer trust this result.  More recently, we obtained H$\alpha$
spectra over four weeks in February 1997 using the 74~inch telescope at
Mt.\ Stromlo.  Those observations were done in collaboration with
I. K. Baldry and M. M. Taylor (Univ.\ Sydney) and analysis is continuing.

During an overlapping period in 1997, \citeone{PBH97} also obtained
observations of Procyon.  The two projects were coordinated under the SONG
program (Stellar Oscillations Network Group\footnote{\tt
http://www.noao.edu/noao/song/}) and we intend to merge the data sets.

\paragraph{Arcturus\hspace{-1ex}} and similar red giants are variable in both
velocity (e.g., \citebare{H+C96} and references within; \citebare{Mer95})
and intensity \citeeg{E+G96}, but the presence of solar-like oscillations
has not yet been established.

\section{Details of the data processing}

Here we describe most of the steps involved in processing a typical data
set.  Step~2 is specific to the equivalent-width method.  The other steps
could apply, at least in part, to other types of observations (Doppler
shift or photometry).
\begin{enumerate}  %\itemsep=0pt \vspace*{-2ex}

\item Preliminary reduction:
	\begin{enumerate}  %\itemsep=0pt \vspace*{-2ex}

	\item Correction for CCD bias by subtracting an average bias frame.

	\item Correction for CCD non-linearity.  Measuring oscillations at
	the ppm level requires that the detector be linear to the level of
	$10^{-3}$ or better.  This is certainly not trivial and our tests
	of different CCDs and controllers often reveal deviations from
	linearity of up to a few per cent.  Unless correction is made for
	these effects, the extra noise will destroy any possibility of
	detecting oscillations.

	\item Correction for pixel-to-pixel variations in CCD sensitivity
	by dividing by an average flat-field exposure.

	\item Subtract sky background, which can be quite substantial
	during twilight.  The background is estimated from the regions at
	each end of the spectrograph slit, above and below the stellar
	spectrum. 

	\item Extraction of one-dimensional spectra.  During this step, the
	seeing in each frame (i.e., the FWHM along the spectrograph slit)
	and the position of the star (i.e., light centroid along the slit)
	should be recorded for possible use in decorrelation (see below).

	\end{enumerate}

\item Measuring equivalent widths:

Achieving high precision requires more than simply fitting a profile.  The
method described here was developed by HK after trying several different
approaches.  By analogy with Str{\o}mgren H$\beta$ photometry, we calculate
the flux in three artificial filters, one centred on the line ($L$) and
the others on the continuum both redward ($R$) and blueward ($B$) of the
line.  For each spectrum, the following steps are followed:
	\begin{enumerate}

	\item With the three filters placed at their nominal positions,
	calculate the three fluxes.\footnote{The flux in a filter is simply
	the total counts in the stellar spectrum after it has been
	multiplied by the filter function.}

	\item Adjust the slope of the spectrum so that $R$ and $B$ are
	equal.  This is done by multiplying the spectrum by a linear ramp.

	\item Re-calculate the filter fluxes and calculate the equivalent
	width: $W = (R-L)/R$.

	\item Move the three filters to a different position and repeat the
	previous steps.  Iterate to find the filter position which
	maximizes the value of $W$.  The other outputs are: position of
	line; height of continuum (from $R$); and slope of continuum.

	\item Repeat the previous steps for four different filter widths.

	\end{enumerate}

\item Initial time series processing:

We now have four times series ($W_1$, $W_2$, $W_3$, $W_4$), one for each
filter width.  Note that the quality of the data, as measured by the local
scatter, usually varies considerably from hour to hour and night to night.
The following procedure is generally applied to each night of data
separately.  
	\begin{enumerate}

	\item Clip each of the four time series to remove outlying points
	(4-$\sigma$ clipping, where $\sigma$ is the local rms scatter).

	\item Calculate weights for each time series.  This involves
	assigning a weight to each data point which is inversely
	proportional to the local rms scatter.

	\item Calculate $\sigma_w$, the {\em weighted\/} rms scatter of
	each time series, using the weights just calculated.  Use this to
	select the best filter width, i.e., the one which minimizes
	$\sigma_w$.  By using a weighted rms scatter, we do not give too
	much importance to the bad segments of the data.  In practice, we
	do not choose one filter width, but rather a weighted combination.
	That is, we choose the powers $a,b,c,d$ to minimize the weighted
	scatter on the time series $W_1^a W_2^b W_3^c W_4^d$, where
	$a+b+c+d=1$.

	\smallskip

	\item[] This step and the subsequent ones rely on the fact that any
	oscillation signal will be much smaller than the rms scatter in the
	time series.  Most of the scatter is due to noise and any method of
	reducing the scatter should be a good thing, although care must be
	taken not to destroy the signal or to introduce a spurious signal.

	\end{enumerate}

\item Decorrelation of time series:

As well as measuring the parameter which is expected to contain the
oscillation signal ($W$), we also monitor extra parameters.  The aim is to
correct for instrumental and other non-stellar effects.  For example, if we
notice that $W$ is correlated with the seeing, we suspect some flaw in our
reduction procedure, since hopefully the stellar oscillation will not know
what is happening in the Earth's atmosphere.  By correlating measured
equivalent widths with seeing variations, one has a chance to remove the
influence of seeing simply by subtracting that part of the signal which
correlates with seeing.  This process of decorrelation, which can be
repeated for other parameters (total light level, position on detector,
slope of continuum, etc.), is very powerful but can also be quite dangerous
if not done with care (see \citebare{GBD91} for a fuller discussion).

Again, the process is done on data from one night at a time.  Performing
decorrelation over shorter intervals runs the risk of moving power around
and creating or destroying signal -- simulations are usual to check these
effects.

\item Calculation of the power spectrum

Once a time series has been extracted, the search for oscillation
frequencies is done by calculating the power spectrum.  The simplest method
is to Fourier transform the time series and take the squared modulus.  The
resulting spectrum shows power as a function of frequency, and a
significant peak in this spectrum implies a periodic signal in the time
series data.  However, the standard Fourier transform treats all data
points as having equal weight.  In reality, data quality can vary
significantly within a data set, due to variable weather conditions or
because data are being combined from different telescopes.  The power
spectrum is very sensitive to bad data points -- the final noise level will
be dominated by the noisiest parts of the time series.  One should
therefore calculate a weighted power spectrum, with each data point being
allocated a statistical weight according to its quality.  In practice, the
power spectrum is calculated as a weighted least-squares fit of sinusoids
\citeeg{FJK95}.  Note that an ordinary Fourier transform is equivalent to
an {\em unweighted\/} least-squares fit.

\end{enumerate}

\section{Conclusion}

In the last few years, the precision in velocity and photometric
measurements has not been significantly improved.  The new equivalent-width
method is far from being fully developed and no confirmation of the claimed
signal in \eBoo{} has been made.  

Space would be a wonderful place to do photometry.  Although the COROT
mission has been selected\footnote{\tt
http://www.astrsp-mrs.fr/www/corotpage.html} and others are being proposed,
for now we will have to continue using ground-based facilities.  It is
important to remember that we are very close to producing noise levels
equal to the solar oscillation signal, and that some stars are expected to
oscillate with higher amplitudes than our own Sun.

\acknowledgments

This work was supported by the Australian Research Council, and by the
Danish National Research Foundation through its establishment of the
Theoretical Astrophysics Center.

\parindent=0pt

\section*{DISCUSSION}

GIBOR BASRI: When using Balmer lines one should be cognizant of the fact
that the line is formed in both the photosphere (wing) and chromosphere
(core) and is a NLTE photoionization-dominated line, and so is not
sensitive to local temperature variations.  Perhaps you are measuring
continuum oscillations?

\medskip

TIM BEDDING: Yes, the oscillations probably are mostly in the continuum.
However, in the presence of scintillation it is impossible to measure
absolute continuum levels, so there is no way to determine whether it is
the line or the continuum that is oscillating.  The only thing which can be
measured with useful precision is the {\em ratio\/} of line to continuum --
in other words, the equivalent width.  To first order, this is insensitive
to scintillation, which is why we choose to observe it.  Note that
spatially resolved observations of the Sun by \citeone{RHD91} and also by
\citeone{KHB97} imply that the Balmer lines are stable and the continuum is
oscillating.

\medskip

BERNARD FOING: What increase of sensitivity do you expect by applying the
temperature method to cross-dispersed echelle spectra, with a proper
weighted combination of equivalent widths of many lines?  If a decisive
advantage, this would allow to use multi-site spectroscopic networks such
as MUSICOS, with a good prospect for solar-type asteroseismology.

\medskip

TIM BEDDING: Yes, there could be a substantial gain in sensitivity.  In
fact, for the AAT observations of \aCenA\ we were able to observe three
orders around both H$\alpha$ and H$\beta$.  In the order next to H$\beta$
there was a strong iron line, which is expected to have a temperature
sensitivity opposite to that of the Balmer lines.  The ratio of the
equivalent widths of iron to H$\beta$ (using suitably weighted powers)
proved to have extremely low scatter.  Essentially, we used the width of
the iron line as a decorrelation parameter and were able to greatly reduce
instrumental effects.  Whether the addition of more (weaker) lines would
give useful improvements is still to be determined.

\medskip

BOB NOYES: Following up on this, simulations carried out by \citeone{NKK96}
suggest that using the {\em entire\/} AFOE spectral range
($\sim$4000\,\AA{} to 6600\,\AA, $R\simeq50,000$) to estimate temperature
change in the photosphere (by comparing changes in line depth at all
wavelengths to predictions of Kurucz models) should allow detection of
individual oscillation modes in Procyen in $\sim$4 nights of observation.
Three attractive features of this approach are: (a)~we can get the data
simultaneously with radial velocity data, so it is `free'; (b)~the
temperature and velocity oscillations are (roughly) in quadrature, hence if
we can detect both, we can use cross-power spectral analysis to strengthen
detection; and (c)~if we can detect both temperature and velocity
oscillations, we can learn more physics from precise phase relations.

\end{document}